\documentstyle[11pt,aaspp4]{article}
\lefthead{K. C. Sahu}
\righthead{Microlensing within the SMC}

\begin{document}

\title{Spectroscopy of MACHO 97-SMC-1: self-lensing within the SMC}

\author{Kailash C. Sahu}
\affil{Space Telescope Science Institute, 3700 San Martin Drive, Baltimore, 
MD~21218\\
ksahu@stsci.edu}

\author{M. S. Sahu}
\affil{NASA/Goddard Space Flight Center, Code 681, Greenbelt, MD~20771\\
and \\
National Optical Astronomy Observatories, 950 N. Cherry Avenue,
Tucson, AZ 8519-4933\\
msahu@panke.gsfc.nasa.gov}

\begin{abstract}
More than a dozen microlensing events have been detected so far 
towards the LMC and 2 towards the SMC. If all the lenses are in the 
Galactic halo, 
both the LMC and the SMC events are expected to have similar time scales.
However, the first event towards the SMC, MACHO 97-SMC-1, had a time scale of 123 days
which is much larger than the typical time scale for the LMC events. 
Since the observed time scale of the SMC event would need the mass of the halo 
lens to be $\sim$3 M$_\odot$, it has been argued earlier that the lens must be within 
the SMC, which we spectroscopically confirm in this paper.

From optical depth estimates, we first show that 
the stars within the SMC play a dominant role as gravitational lenses 
and can fully account for the observed microlensing events, mainly due 
to its large physical depth. We also show that if the lenses are within 
the Magellanic Clouds, then the SMC events should be longer in duration 
than the LMC events, a fact that is consistent with the observations. 
The time scale of the event implies that the mass of the lens is 
$^>$\hskip-0.25cm $_\sim$ \ 2 M$\odot$ if it is in the Milky Way disk or 
halo, in which case
the lens is expected to be bright and should reveal itself in the spectrum.
Here, we present an optical spectrum of MACHO 97-SMC-1 obtained in May 1997
which shows that the source is a main-sequence B star. There is no trace 
of any contribution from the lens which demonstrates that
the lens is not in the Milky Way disk or halo, but is a low-mass
star within the SMC.

It is worth noting here that MACHO SMC-98-1 is the only {\it other}
observed event towards the SMC. This  was a binary lens event where the 
caustic crossing time-scale as observed by PLANET, MACHO, EROS and OGLE  
collaborations, suggests that the lens is within the SMC.
Furthermore, the only LMC event where the location of the lens is known
is the binary lens event MACHO LMC-9, for which the lens is also
within the LMC. Thus, {\it all} the 3 microlensing events towards the
Magellanic Clouds for which we have independent knowledge of the location 
of the lenses are due to self-lensing within the Magellanic Clouds.

\end{abstract}


\section{Introduction}

In five years of monitoring millions of stars towards the Large and Small 
Magellanic Clouds, more than a dozen microlensing 
events have been detected towards the LMC and 2 towards the SMC
(Alcock et al. 1997a; Afonso et al. 1998). 
The lenses have been
hypothesized to be either in the Milky Way disk (Gould, Miralda-Escud\'e 
and Bahcall 1994; Evans et al. 
1998), halo (Alcock et al. 1997a),  the Magellanic Stream (Zhao, 1998),
or the LMC disk itself (Sahu, 1994 a,b; Wu, 1994).
If the lenses are in the Galactic halo, the implied mass of the lenses is of the
order of 0.5 M$_\odot$.
This poses problems for the halo origin of the lenses 
since such lenses would
be bright and numerous, and hence be directly observable if they are
stellar objects, which is
inconsistent with deep HST (Flynn, Gould and Bahcall, 1994) and ground-based 
(Hu et al. 1994) observations. One way to distinguish between whether the
lenses are in the Milky Way disk/halo or in the Magellanic Clouds is the
following: if the lenses are within the Galactic disk or the halo,
the time scales for the LMC and the SMC events should be similar 
though not identical if the halo is highly flattened (Sackett and Gould, 1993). 
On the other hand, if the lenses are within the 
Magellanic Clouds, the time scales for the SMC events, as we show in
this paper, are expected to be much longer. 
Here we calculate the self-lensing optical depth
taking the structure of the SMC into account and show that the SMC stars
play a dominant role as gravitational lenses.
In the case of SMC MACHO 97-SMC-1 (Alcock et al. 1997b) the time scale  has been
used earlier to infer that the lens is likely within the SMC 
(Palanque-Delabrouille et al. 1997). We present spectroscopic observations
which confirm that the lens is indeed within the SMC. 
\section{Optical Depth for self-lensing within the SMC}
The simple microlensing light curve has a degeneracy between
the mass and the perpendicular velocity and is hence insufficient 
to provide information about the location of the lens.
Thus, one is forced to resort to statistical means to determine the
location of the lenses. In such a statistical determination,
it is important to calculate the contribution of stars which
are known to exist within the Milky Way disk and within the SMC.
The contribution of the Galactic stars to the microlensing 
optical depth is not large and has been estimated by
Gould, Miralda-Escud\'e and Bahcall (1994). The
contribution of the SMC stars to the microlensing optical depth
has recently been estimated by (Palanque-Delabrouille et al. 1997)
taking a smooth density distribution within the SMC. 
Here we first recalculate the optical depth with a different approach
 taking the observed structure
and the extent of the SMC into account where, in the smooth density 
distribution case, the results are consistent with that of 
Palanque-Delabrouille et al.
We also show the optical depth can be larger if the SMC has separate 
components as observed.

To calculate the optical depth by stars within the SMC, one may like 
to use the observed velocity dispersion
where, under some simplified assumptions, the microlensing optical depth
can be written as  (Paczy\'nski, 1996) 
$\tau  \simeq   <v^2>/c^2$, 
with the resultant microlensing optical depth 
$\tau =  10^{-8}$ (for $v$ = 30 km s$^{-1}$).
 However, the velocity dispersion can be used only if the system is 
self-gravitating and the density is uniform so that the virial theorem 
or Jean's equation  (Paczy\'nski, 1996; Gould, 1995) can be applied. The 
SMC is  known to be far from such a system; and as described below,
it has 4 kinematically
distinct structures and has a large depth which ranges from 5 to 20 kpc 
depending on the region under study. Clearly, we can not use the velocity 
dispersion but must use the luminosity and structure of the SMC instead, 
to estimate the optical depth. 

{\subsection{Structure of the SMC}}

The mean distance modulus to the SMC is 18.9 as derived recently
(Barnes, 1983; Sebo and Wood, 1994; Walker, 1998) which corresponds to 60 kpc.
From a study
of A, B and O supergiants a distance spread of $\sim$ 7 kpc within the SMC was
first derived by Azzopardi (1982); a more recent study however  shows
that the  distance spread is $>$ 10 kpc (Hatzidimitriou and Cannon, 1993).
From a study of IR period-luminosity relation of 161 Cepheids distributed over
a wide area of the SMC, Mathewson et al. (1986) found that the 
depth of the SMC is $>$ 20 kpc with the maximum concentration of
Cepheids at $\sim$ 59 kpc. Martin et al. (1989) conclude
that most of the young SMC stars lie within a depth of $<$ 10 kpc, which 
roughly corresponds to the lateral diameter of the SMC. 

A vast amount of data are available which show that the kinematical 
structure of the SMC is also very complex. The extensive HI observations 
of McGee and Newton (1986) show that the SMC has 4 principal and overlapping  structures at velocities
V$_{hel}$ = 114, 133, 167 and 192 km s$^{-1}$, each component having a velocity dispersion
of $\sim$ 25 km s$^{-1}$ (see Fig. 2 of McGee and Newton, 1986). 97\% of the 
506 line profiles show multiple peaks.
Torres and Carranza (1987) show that radial velocities of
supergiants and interstellar calcium absorption features 
correlate well
with the HI velocities. An HI aperture synthesis mosaic of the SMC
has been recently published by Staveley-Smith et al. (1997) which shows all
these structures. 

{\subsection{Microlensing Optical Depth}}

The microlensing
optical depth due to self-lensing within the SMC can be written as
(Sahu, 1994 a,b)
$$\tau = {1\over{N_{tot}}}\int_0^d{N_{obs}(z) A_f(z)  dz} \eqno (1) $$
where N$_{tot}$ is the total number of stars being monitored,
and N$_{obs}(z)$ is the number of stars at a layer which
is at a depth $z$ within the SMC, $A_f(z)$ is the
fractional area covered by the Einstein rings of all  the individual
stars lying in the front of this layer, and $d$ is the total depth of the
SMC. Since the internal extinction 
within the SMC is low (Westerlund, 1997), we neglect the effect of extinction 
on the number of observed stars at different depths within 
the SMC, in which case the optical depth can be expressed as
$$\tau = \int_0^d{A_f(z)  dz} = \int_0^d{\pi R_E^2 n(z) dz} \eqno (2) $$
where $n(z)$ is the stellar number density at depth $z$. 
Since $n = \rho /m$
and $R_E^2 \propto m$ where $\rho$ is the stellar mass density and $m$ is
the mass of the star, 
we substitute 
 $\rho = {M  \over {A \times  d}} $
so that 
$$\tau  \simeq {2 \pi G M  d \over  c^2 A} \eqno (3)$$
where $M$ and $A$ are the total mass and the projected area on the sky of 
the SMC, respectively.
 Using M = 2 $\times$ 10$^9$ M$_\odot$   and Area = 15 kpc$^2$ (Westerlund,
1997), 
$\tau = 2.0 \times 10^{-7} {d\over{5 kpc}}$.

Note that $\tau \propto d$, so the
large depth of the SMC makes its self-lensing optical depth larger.
On the other hand, if the SMC has a layered structure as
described above, the optical depth can be higher. 
For simplicity, let us assume that the SMC has two layers with equal
mass separated by a distance $h$. If $h >> d$, the optical depth 
can be expressed as
$$\tau  \simeq {2 \pi G M  h \over  c^2 A} +
{2\pi G \over c^2}\int_0^{d_1}{\rho (z) z dz} +
{2\pi G \over c^2}\int_0^{d_2}{\rho (z) z dz} 
\eqno (4)$$
\noindent where $h$ is the separation between the two disks and $d_1$ and
$d_2$ are the depths of individual disks. Here, the first term is the 
contribution from the lensing of the background layer by the foreground,
the second and third terms are the self-lensing optical depths in the
two individual disks. Assuming the 
depths of the two disks to be equal ($d$) and assuming the
density to be equal in both disks, 
the optical depth can be expressed as
$$\tau  \simeq {2 \pi G M  (h+2d) \over  c^2 A} 
 \sim 2.4 \times 10^{-7} {h\over{5 kpc}}  \eqno (5)$$

Considering the fact that the density is non-uniform, and the
surface luminosity varies by more than a factor of 4 across the SMC,
the full range of the self-lensing optical depth for the SMC is
$$1.0 \times 10^{-7} < \tau < 5.0 \times 10^{-7} \eqno (6)$$

The exact observed microlensing optical depth is uncertain 
(Alcock et al. 1997b), but is roughly in this range.

\subsection{Contribution from the bar}

The SMC has a noticeable bar with a size of $\sim$ 2.5 $\times$ 1 kpc.
The surface luminosity in this area is about 5 times higher than the
average surface luminosity of the rest of the
SMC and hence one may naively expect that the microlensing optical depth 
would be about 5 times higher in this region. However, as
seen from Eq 4, the self-lensing optical depth in such a case is
directly proportional to the depth and can be
expressed as 
$$\tau_{bar} \sim \tau_{disk} \times {S_{bar}\over{S_{disk}}} \times 
{d_{bar}\over{d_{disk}}} \eqno (7)$$
where $S$ refers to the surface luminosity and $d$ refers to the depth.
Thus the contribution of the surface luminosity 
may be more than compensated by the effect of the depth.
The effect of extinction may also be important in the region
of the bar, which decreases the optical depth
(for details, see Sahu, 1994b). Furthermore, the crowding effect
is expected to be more severe in the region of the bar, which 
increases the number of monitored stars but decreases the efficiency
of detection. Without the full knowledge of the extinction 
and crowding in this region,
it is hard to do a quantitative estimate of the distribution of
optical depth in the bar and its surroundings. 
It is reasonable to expect, however, that there may be slight enhancement 
of the optical depth in the outer region of the bar where the effect of
crowding and extinction is less severe and hence the background stars can 
be monitored. It is worth noting that both the microlensing events 
observed in the SMC are about 1 degree away from the center of the bar. 

\section{Time scales}

The Einstein ring crossing time $t_E$ is given by
$$t_E = {R_E\over{V_\perp}} = {\sqrt{4GMD}\over{cV_\perp}}; \ 
D = {D_L D_{LS}\over{D_s}} \eqno (8)$$
where $V_\perp$ is the source-lens proper motion at the lens plane, 
M is the  the mass of the lens,  D$_L$ is the the distance to the lens,
D$_{LS}$ is the the distance from the lens to the source, and
D$_S$ is the distance to the source. 
So the time scales in different cases can be written as

$$t_{halo} = 37 {\sqrt{{M\over{0.2 M_\odot}} {D_L\over{15 kpc}}}}
{200kms^{-1}\over{V_{\perp}}} days \eqno (9)$$

$$t_{SMC} = 105 {\sqrt{{M\over{0.2 M_\odot}} {D\over{5 kpc}}}}
{45kms^{-1}\over{V_{\perp}}} days \eqno (10)$$
 
$$t_{LMC \ or \ disk} = 33 {\sqrt{{M\over{0.2 M_\odot}} {D\over{200 pc}}}}
{30kms^{-1}\over{V_{\perp}}} days \eqno (11)$$
 
\noindent where $t_{halo}$, $t_{SMC}$, $t_{LMC}$, and $t_{disk}$
are the time scales for lensing by halo
objects,  self-lensing within the SMC,  self-lensing within
the LMC, and objects in the Milky Way disk, respectively. 
Note that we have used
a typical velocity of 45 km s$^{-1}$ within the SMC,
in accordance with the discussion in section 2.1. 

Thus the time scales could be a clear distinguishing feature.
If the microlensing events  are caused by the halo lenses, 
then the time scales for both the LMC and the 
SMC events should be similar. If they are caused by self-lensing, the
SMC events should be much longer.

\section{The first SMC microlensing event MACHO 97-SMC-1}

From the color and luminosity of the source alone, the source 
in the case of MACHO 97-SMC-1 is expected to be a
B-type star in the SMC. The light curve shows a periodic variation with a 
peak-to-peak amplitude of about 0.2 magnitude and a periodicity of 
about 5 days, both inside and outside the microlensing period (Udalski et al.
1997). This has been interpreted as due to the source being an ellipsoidal 
variable where the individual components of the binary are tidally 
disrupted (Paczy\'nski 1997). This event had a peak amplification
of 2.1 and a time scale $t_E$ of 123 days. 

\subsection{Spectroscopic observations and implications}

We used the 
EFOSC (ESO Faint Object Spectrographic Camera) at the
ESO 3.6m telescope at La Silla, Chile on 30th May 1997,
to obtain a spectrum of the source just after the event was 
over. The camera has both imaging and spectroscopic capabilities. 
First, a direct image was obtained using the 512 $\times$ 512 pixel 
CCD with a field of view of 5.2$\arcmin \times$ 5.2$\arcmin$.
The first spectra were obtained with the B150 grating, with an 1.5$\arcsec$
slit oriented along east-west. This covers the wavelength range 3900 - 5300 
\AA, with a resolution of about 5\AA. Another spectrum was obtained 
using the O150 grating which covers the wavelength range 5200 - 6800 \AA,
with the same resolution. The seeing was about 1$\arcsec$, so the 
contribution from the nearby `blending star', which affected some photometric
observations of the monitoring programs, is minimal.

\begin{figure}
\plotone{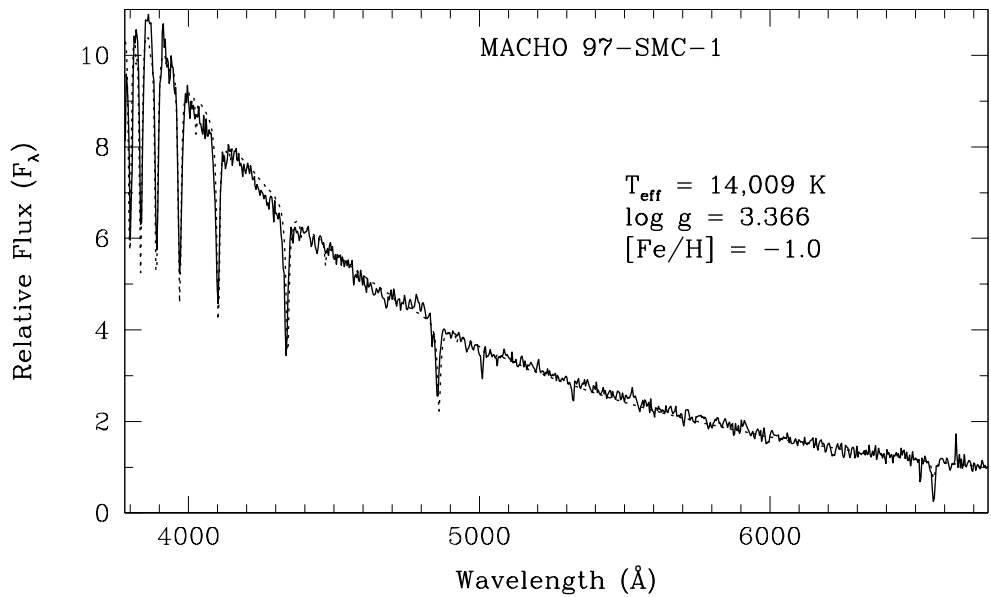}
\caption{Observed spectrum of MACHO 97-SMC-1 taken on  30th May 1997,
just after the source had crossed the Einstein ring
radius of the lens (solid curve), along with the
best fit stellar model spectrum (dashed line). The contribution from the lens
is negligible which, combined with the microlensing time scale, 
implies that the lens must be within the SMC (see $\S 4.1$ for details).
\label{fig1}}
\end{figure}

The data were reduced using the standard spectroscopic reduction
procedures available in the software packages MIDAS and IRAF.
The sky was taken from regions above and below  the spectrum for a good sky 
subtraction and 
the standard star LTT 9239 was used for flux calibration. The resulting 
final spectra for the source
is shown in Fig. 1. The figure also shows a best-fit stellar model 
to the spectrum. A metallicity of $-1.5$ corresponding 
to the SMC used for the fit and the effective stellar temperature and 
surface gravity
were determined simultaneously from a grid of synthetic spectra (Saffer et al. 
1994). The derived effective stellar temperature is 14000 K and 
$log\  g$ =3.66, which corresponds to a main sequence B star. The resolution 
of the spectrum does not allow checking for blending of lines
from two components, so from this spectrum alone it is difficult
to confirm if the source is indeed an ellipsoidal variable star.
The continuum is fit perfectly well
with a single star spectrum, without any need for an extra component.
There are, however, very weak additional lines
around 5018 and 5316  \AA \ which may be FeII lines 
(they are unlikely to be He lines since other He lines are not observed)
indicating that the metallicity may be slightly higher, or the 
`blending star' may have a small contribution to the absorption lines.
The contribution from the lensing star is clearly
insignificant to the continuum. 

The spectrum of the source taken at base-line (so that the contribution
from the lens is maximum) gives a good opportunity
to check  the contribution of the lens (Kane 
and Sahu, 1998;  also see Mao, Reetz and Lennon, 1998). 
From Eq. 9, the mass of the lens is $\sim$ 2.9 M$_\odot$
if the lens is in the halo. A normal star of such high mass should have 
significant contribution to the spectrum which is not the case.
The lens cannot be in the Milky Way disk for several reasons: first, 
the probability of lensing by a disk object is very low. Secondly, the 
estimated mass of a disk lens is $\sim$ 1.9 M$_\odot$. A star of this mass at
300 pc would have $m_v \sim 9.5$, which is clearly not the case. 
However, the mass distribution for disk lenses is rather wide,
and the lens could still be a low mass star. But a lens in the local disk
should have shown the ``parallax effect" (Alcock et al. 1995), particularly
since the time scale is large, which is not observed.
The estimated mass for an SMC lens is $\sim$0.2 M$_\odot$,
in which case the luminosity of the lens is too low
to have any contribution to the spectrum, which is consistent with
the observations. This demonstrates that the lens 
is not in the Milky Way disk or halo, but is within the SMC.

\section{Conclusions}

Taking the structure and extent of
the SMC into account, we have shown that the 
stars within the SMC play a strong role as lenses and
can indeed account for all the observed microlensing 
events towards the SMC. Furthermore, if the lenses are in the halo,
the time scales for the LMC and the SMC events should be similar.
On the other hand, if the lenses are
within the Magellanic Clouds, then the time scales for the SMC 
events should be larger than the LMC events, which is consistent with 
the observations.

The observed time scale of MACHO 97-SMC-1 implies that the lens mass
is $^>$\hskip-0.25cm $_\sim$ \ 2 M$\odot$ if the lensing is caused by an 
object in the halo
or the disk, in which case the spectrum should show the contribution
from the lens. The fact that there is no contribution of the lens 
in the observed spectrum further supports the conclusion
that the lens is within the SMC.

The only other lensing event towards the SMC,
MACHO SMC-98-1, also is a binary lens event where
the caustic crossing time scales derived from the observations by PLANET
(Albrow et al. 1998), MACHO (Alcock et el. 1998), EROS 
(Afonso et al. 1998) and OGLE (Udalski et al. 1998)
collaborations imply that the lens in this case too, is within the SMC.
Thus, we have some independent knowledge of the
location of the lenses in both cases of observed
microlensing events towards the SMC, and they
{\it both} are due to self-lensing within the SMC.
Towards the LMC, there is only one binary event, MACHO-LMC-9,
for which we have independent knowledge of the location of the lens
(Bennett et al. 1996) where the lens is inferred to be within the LMC.
Thus, out of all the events observed so far towards the
LMC and the SMC, there are a total of 3 events for which we 
have some independent knowledge of  the location
of the lenses: 2 towards the SMC as stated above, and one towards the 
LMC (MACHO-LMC-9).
In all cases without exception, the lens is within the
Magellanic Clouds, which supports the earlier suggestion
by Sahu (1994a,b) that the stars within the Magellanic Clouds play
a dominant role as gravitational lenses.

\acknowledgements{We thank Rex Saffer for his ready help
with the use of his spectral classification code.} 

{\parindent=0pt
\parskip=0pt
{\bf{References:}}

Afonso, C., et al. (EROS Collaboration), 1998, astro-ph/9806380\\
Albrow, M., et al. (PLANET Collaboration), 1998, astro-ph/9807086\\
Alcock, C. et al. 1998, astro-ph/9607163\\
Alcock, C. et al. 1997a, ApJ, 486, 697\\
Alcock, C. et al. 1997b, ApJ, 491, L11\\
Alcock, C. et al. 1995, ApJ, 454, L125\\
Azzopardi, M.  1982, Comptes Rendus sur les Journees de Strasbourg, 
Observatoire de Strasbourg, p20\\
Barnes III T. G., Moffett, T.J., Gieren, W.P. 1993, ApJ, 405, L51\\
Bennett,D.,P. et al. 1996, in Nucl. Phys. B. (Proc. Suppl.), 51B, 131\\
Evans, N. W.,  Gyuk, G., Turner, M. S.,
  Binney, J. 1998, ApJ, 501, L45\\
Flynn, C., Gould, A., Bahcall, J.N. 1996, ApJ., 466, L55\\ 
Gould, A., Miralda-Escud\'e, J., Bahcall, J. 1994, ApJ, 423, L105\\
Gould, A., Bahcall, J., Flynn, C. 1996, ApJ, { 465}, 759\\
Gould, A. 1995, ApJ, 441, 77\\
Hatzidimitriou, D., and Cannon, R. 1993, New Aspects of Magellanic
Cloud Research, p17\\
Hu, E. M., Huang, J. S., Gilmore, G., Cowie, L. L. 1994, Nature, 371, 493\\ 
Kane, S., Sahu, K.C. 1998, in preparation.\\ 
Mao, S., Reetz, J., \& Lennon, D.J. 1998, A\&A, 338, 56\\
Martin, N., Maurice, E., Lequeux, J. 1989, A\& A, 215, 219\\
Mathewson, D.S., Ford, V.L. Viswanathan, N. 1986, ApJ, 301, 664\\
McGee, R.X., Newton, L.M. 1986, Proc. ASA, 4, 305\\
Palanque-Delabrouille, N., et al. (EROS collab) 1997, Astron. Astrophys. 332, 1\\
Paczy{\'n}ski, B. 1996, Acta Astron. 46, 291\\
Paczy\'nski, B. 1997,  astro-ph/9711007\\
Sackett, P., Gould, A. 1993, ApJ, 419, 648\\
Saffer, R. A., Bergeron, P., Koester, D., Liebert, J. 1994,
ApJ., 432, 351\\
Sahu, K.C. 1994a, Nature, 370, 265\\
Sahu, K.C. 1994b, Pub. Astr. Soc. Pac., { 106}, 942\\
Sebo, K.M., Wood, P.R., 1994, AJ, 108, 932\\
Staveley-Smith, L, Sault, R.J., McConnell, D., Kesteven, M.J.
1997, MNRAS, 289, 225\\
Torres, G., Carranza, G.J. 1987, MNRAS, 226, 513\\
Udalski, A. et al. 1997, Acta. Astron. 47, 431\\
Udalski, A. et al. 1998,  astro-ph/9808077\\
Walker, A.R. 1998, astro-ph/9808336\\
Westerlund, B.E. 1997, The Magellanic Clouds, Cambridge Univ. Press,
p28-34\\
Wu, X-P. 1994, ApJ., { 435}, 66\\
Zhao, H.S. 1998, MNRAS, 294, 139\\

\end{document}